\documentclass[reprint,floatfix,showpacs,twocolumn,10pt]{revtex4-1}

\usepackage{graphicx}
\usepackage{amsmath}
\usepackage{amsfonts}
\usepackage{amssymb}
\usepackage{nameref}
\usepackage{amsthm}
\usepackage[colorlinks]{hyperref}
\usepackage{color}
\usepackage[utf8]{inputenc}

\newcommand{\bra}[1]{\ensuremath{\left\langle#1\right|}}
\newcommand{\ket}[1]{\ensuremath{\left|#1\right\rangle}}
\newcommand{\braket}[2]{\ensuremath{\left\langle#1 \vphantom{#2}\right| \left. \!\! #2 \vphantom{#1}\right\rangle}}
\newcommand{\matrixel}[3]{\ensuremath{\left\langle #1 \vphantom{#2#3} \right| #2 \left| #3 \vphantom{#1#2} \right\rangle}}
\newcommand{\ketbra}[2]{\ket{#1}\!\!\bra{#2}}

\newtheorem{theorem}{Theorem}

\graphicspath{ {./Figures/} }

\DeclareMathOperator{\tr}{tr}
\DeclareMathOperator{\real}{Re}

\newcommand{\sop}[1]{\tilde{\mathcal{O}}^{(#1)}_{D_4}}
\newcommand{\eqnref}[1]{Eq.~(\ref{#1})}

\newcommand{\onehalf}{\frac{1}{2}}
\newcommand{\halfpi}{\frac{\pi}{2}}
\newcommand{\RgEi}{\tilde{\mathcal{E}}_I}
\newcommand{\tildeXi}{\tilde{\xi}}

\begin{document}

\title{Resource quality of a symmetry-protected topologically ordered phase \\
       for quantum computation}

\author{Jacob Miller}
\email{jmilla@unm.edu}
\author{Akimasa Miyake}
\email{amiyake@unm.edu}
\affiliation{Center for Quantum Information and Control, Department of Physics and Astronomy, University of New Mexico, Albuquerque, NM 87131, USA}

\begin{abstract}

We investigate entanglement naturally present in the 1D topologically ordered phase protected with the on-site symmetry group of an octahedron as a potential resource for teleportation-based quantum computation. We show that, as long as certain characteristic lengths are finite, all its ground states have the capability to implement any unit-fidelity one-qubit gate operation asymptotically as a key computational building block. This feature is intrinsic to the entire phase, in that perfect gate fidelity coincides with perfect string order parameters under a state-insensitive renormalization procedure. Our approach may pave the way toward a novel program to classify quantum many-body systems based on their operational use for quantum information processing.

\end{abstract}

\maketitle

\emph{Introduction.---}Entanglement is ubiquitous in quantum many-body systems, and its complexity has drawn attention from interdisciplinary research fields, such as condensed-matter physics \cite{li2008entanglement, pollmann2010entanglement, chen2010local, deChiara2012entanglement}, quantum information processing (QIP) \cite{raussendorf2001one, raussendorf2003measurement, vidal2003efficient}, and quantum simulation of quantum many-body systems \cite{cirac2012goals, korenblit2012quantum, georgescu2014quantum, eisert2014quantum, cohen2014proposal}. A primary example is exotic ground states of topologically ordered phases \cite{wen2007quantum, hasan2010topological, qi2011topological}, which arise from underlying nonlocal entanglement. It is widely known that braiding their excitations, known as anyons, could be used for topological quantum computation \cite{kitaev2003fault}, and their intrinsic insensitivity against local noise could be used for quantum error correction \cite{kitaev2003fault, kitaev2010topological}. Many-body entanglement can be harnessed in a more direct way, and certain many-body states like 2D cluster states \cite{briegel2001persistent} and certain tensor network states \cite{verstraete2004valence, gross2007novel, gross2007measurement, cai2010universal, miyake2011quantum, wei2011affleck, li2011thermal} are quantum resources for measurement-based (or teleportation-based) quantum computation, in that universal quantum computation can be implemented on these states using only single-spin measurements.

Having in hand a long list of many-body entanglement useful for QIP, however, one may wonder ``Is such computational usefulness robust in the same way that collective phenomena of quantum many-body systems do not depend on their microscopic details?'' Phrased differently, ``Can we define quantum phases useful for certain QIP tasks in the same way we define phase diagrams in condensed matter physics, which are typically characterized by order parameters?'' There have been several attempts \cite{browne2008phase, doherty2009identifying, barrett2009transitions, skrovseth2009phase, bartlett2010quantum, darmawan2012measurement, else2012symmetrynewjphys, fujii2012topologically, fujii2013measurement} to answer this affirmatively, but they unfortunately, with a few exceptions \cite{bartlett2010quantum}, were largely based on a limited class of states, using rather artificial Hamiltonians from a condensed matter physics perspective.

Here we tackle this challenge using the 1D counterpart of topologically ordered phases as a key building block for measurement-based quantum computation, taking advantage of recent characterizations of symmetry protected topologically ordered (SPTO) phases \cite{chen2011classification, gu2009tensor, schuch2011classifying, pollmann2012symmetry}. By inventing a physically-feasible renormalization procedure which extracts the robust, macroscopic features common among ground states within a phase, we prove that all the ground states in the 1D SPTO phase corresponding to octahedral on-site symmetry can be used to implement any one-qubit operations perfectly, as long as certain conditions on characteristic length scales are met. The leverage of a discrete symmetry is somehow reminiscent of magic states and their distillation \cite{bravyi2005universal} in the context of fault-tolerant, universal quantum computation. Furthermore, we show that the gate fidelity, which is a typical measure of resource quality in QIP, can be interpreted as an ``operationally-motivated'' order parameter of the phase, because it detects critical points of the phase in the same way as the conventional string order parameter widely used in condensed matter physics.  As a whole, our results constitute the first solid evidence for quantum computationally useful phases of matter.

\emph{Matrix product states and 1D symmetry-protected topological orders.---}The matrix product state (MPS) formalism \cite{fannes1992finitely, vidal2003efficient, perez2006matrix} is an efficient means of describing the correlations in one-dimensional spin chains. A MPS description is given by associating a matrix, $A_i$, to every vector $\ket{i}$ of a single-spin basis $\left\{ \ket{i} \right\}_{i=1}^d$. The amplitude associated with a basis vector $\ket{i_1 i_2 \ldots i_n}$ is then given by
\begin{equation}
  \label{eq:Mps2Amp}
   \braket{i_1 i_2 \ldots i_n}{\psi}= \tr \left( A_{i_1} A_{i_2} \cdots A_{i_n} \right).
\end{equation}
The correlation length of our MPS is denoted by $\xi$, and our MPS is short-range correlated if $\xi$ is finite.

In the presence of an on-site symmetry group $G$, $G$-invariant MPS's form distinct symmetry protected topological ordered (SPTO) phases, a classification of which was given in Refs.~\cite{chen2011classification,schuch2011classifying}. Any transition between SPTO phases must be accompanied by either the introduction of long-range correlations or the breaking of on-site symmetry. This makes SPTO phase a robust property of many-body systems in the presence of symmetry. The group of $\pi$ rotations around the $x$, $y$, and $z$ axes, $D_2 \simeq Z_2 \times Z_2$, defines two quantum phases, the trivial phase and the $D_2$ SPTO phase. The archetypical member of the $D_2$ SPTO phase is the Affleck-Kennedy-Lieb-Tasaki (AKLT) state \cite{affleck1988valence}, whose MPS matrices are $A_\mu = \sigma_\mu$. $\mu$ labels the vectors in the spin-1 Pauli basis $\{ \ket{\mu} \}_{\mu=1}^3$, defined by $S_\mu^{(1)} \! \ket{\mu} = 0$, with $S_\mu^{(1)}$ the spin-1 angular momentum operators. The $\sigma_\mu$ are the standard spin-$\onehalf$ Pauli operators.

Measurement-based quantum computation (MQC) \cite{raussendorf2001one,raussendorf2003measurement} is a convenient setting for quantum computation where the quantum nature of computation comes from the entanglement of an initial resource state. Through a sequence of single-spin measurements, an MQC protocol harnesses this entanglement to implement a quantum algorithm. In this paper, we focus on one-dimensional resource states, which are an essential building block for constructing universal resource states for quantum computation. As an illustration, we examine an MQC protocol utilizing the AKLT state \cite{brennen2008measurement}. If we measure a spin in our AKLT chain and obtain an outcome $\ket{\psi_k} = \sum_{\mu=1}^3 \psi_{k,\mu} \ket{\mu}$, then this results in an operator
\begin{equation}
  \label{eq:AKLTMeasurement}
  A[\psi_k] = \sum_{\mu=1}^3 \psi_{k,\mu}^* A_\mu = \sum_{\mu=1}^3 \psi_{k,\mu}^* \sigma_\mu.
\end{equation}
\noindent If we wish to implement a rotation by $\Theta$ around the $z$ axis, $U_\Theta = \exp(-i \frac{\Theta}{2} \sigma_z)$, a measurement outcome of $\ket{\psi_{z,\Theta}} = \cos\!\left(\frac{\Theta}{2}\right) \ket{x} - \sin\!\left(\frac{\Theta}{2}\right) \ket{y}$ will suffice, since
\begin{equation}
  \label{eq:AKLTRotation}
  A[\psi_{z,\Theta}] = \sigma_x \left[ \cos\!\left(\frac{\Theta}{2}\right) I - i \sin\!\left(\frac{\Theta}{2}\right) \sigma_z \right]
\end{equation}
\noindent is indeed what we wanted, up to the $\sigma_x$ term. This additional term is referred to as a byproduct operator, and can be dealt with as long as we maintain a record of the operator (See \cite{raussendorf2003measurement} for details).

\emph{Motivations of our work.---}The above protocol characterizes one point within the $D_2$ SPTO phase, namely the AKLT state, as a resource state capable of generating arbitrary one-qubit operations. As stated in the Introduction, to explore whether such a resource characterization can be extended to the rest of the $D_2$ SPTO phase, we wish to invent a state-insensitive MQC protocol, in that an identical computation should be generated despite microscopic differences of ground states. An initiative along this direction was taken in \cite{bartlett2010quantum}, where all ground states of the 1D $SO_3$-invariant Haldane phase (or the so-called bilinear-biquadratic Hamiltonians) were studied using DMRG calculations. The perfect resource quality of these states for arbitrary single-qubit operations was demonstrated heuristically using a renormalization argument mapping any ground state towards the AKLT state. Later, Else \emph{et.\ al.}\ \cite{else2012symmetryprl} developed an algebraic characterization of the $D_2$ SPTO phase, which includes the $SO_3$-invariant Haldane phase, showing that any state within this phase can be used to implement a state-insensitive qubit teleportation operation. They obtain this result by showing that \cite{Footnote1} for any spin-1 MPS within the $D_2$ SPTO phase, the component matrices associated with that state's MPS have the form
\begin{equation}
  \label{eq:ProtJunk}
  A_\mu = \sigma_\mu \otimes a_\mu.
\end{equation}
\noindent The Hilbert spaces on the left and right side of the tensor product in \eqnref{eq:ProtJunk} are called the protected space and the junk space, respectively. While the details of the junk operators, $a_\mu$, vary from state to state, the structure of the protected space is common everywhere throughout the $D_2$ SPTO phase. Thus, if we measure our resource state in the Pauli basis, we will always end up teleporting the state of the protected space. In retrospect, this feature was first observed for certain ground states of the $D_2$ SPTO phase, like in the spin-1 XXZ Heisenberg model, as its so-called localizable entanglement diverges, and can thus be used to implement the identity channel \cite{popp2005analytic, venuti2005analytic}.

However, a simple argument given by Else \emph{et.\ al.}\ \cite{else2012symmetryprl} suggests that the resource characterization of the $D_2$ SPTO phase is limited to the identity channel (namely teleportation). If we perform some non-Pauli measurement, such as that in \eqnref{eq:AKLTRotation}, we end up applying the operation
\begin{eqnarray}
  \label{eq:ProtJunkEntanglement}
  A[\psi_{z,\halfpi}] &=& \cos\!\left( \frac{\Theta}{2} \right) I \otimes a_x - i \sin\!\left( \frac{\Theta}{2} \right) \sigma_z \otimes a_y \nonumber \\
                        &\neq& \left[ \cos\!\left(\frac{\Theta}{2}\right)I - i \sin\!\left(\frac{\Theta}{2}\right)\sigma_z \right] \otimes a_{x}.
\end{eqnarray}
\noindent Because $a_x \neq a_y$ for arbitrary states, this operation generally won't have a well-defined effect on the protected space, and thus doesn't implement a state-insensitive unitary rotation within the $D_2$ SPTO phase.

\emph{Main Results.---}Now we focus on MPS's invariant under on-site octahedral symmetry. This group can be generated by $\halfpi$ rotations around the $x$ and $z$ axes of the octahedron, and is actually isomorphic to the symmetric group of degree 4, $S_4$. Since the $\pi$ rotations in $S_4$ generate the group $D_2$, any state with $S_4$ symmetry also has $D_2$ symmetry. It can be shown that the classification of SPTO phases for on-site $S_4$ symmetry is identical to the case of $D_2$, and consequently, any MPS in the $S_4$ SPTO phase is automatically in the $D_2$ SPTO phase. This makes \eqnref{eq:ProtJunk} applicable also to states in the $S_4$ SPTO phase, but the larger symmetry of $S_4$ imposes finer constraints on MPS's in the $S_4$ SPTO phase. We emphasize that this abstract characterization of SPTO phases is useful for making general statements, like the following two theorems, without specifying a system Hamiltonian or other microscopic details (although one could define a formal, local Hamiltonian for every MPS).

We study this $S_4$ SPTO phase by means of an operational ``renormalization'' protocol called $z$-buffering, which extracts macroscopic features common among ground states within the phase. This protocol, shown in Figure~\ref{fig:Schematic}, consists of sequential {\it single-spin} measurements, with postselection for a desired measurement outcome which depends on the type of rotation we wish to implement. We first select a site, the computational site, which will eventually be used to generate the desired unitary rotation. Pauli measurements are then performed on the $m$ sites on each side of this site. If we want to implement a $z$-axis rotation using the computational site, we postselect for the all-$\ket{z}$ outcome on these $2 m$ buffering sites, a process called $z$-buffering. Similarly for $x$-axis rotations, $x$-buffering is utilized by postselecting the all-$\ket{x}$ outcome. The ability to perform $z$ and $x$-axis rotations is all we need, since any single-qubit unitary gate can then be constructed using Euler angles.

If our desired outcome isn't obtained, we just measure the computational site in the Pauli basis and repeat this process on the next part of our spin chain, the state of our protected space simply being teleported by this undesired measurement outcome. Note that the probability of postselection is accounted for as overhead in the chain length, but this does not qualitatively change the resource quality (and its complexity), as long as it is finite. On the other hand, if our postselection succeeds, then the remaining computational state is renormalized by an amount depending on the ratio of $m$ to a characteristic length scale, called the $z$-correlation length $\zeta_z$, which governs this RG flow for each state. When $\zeta_z$ is finite, this RG flow generally terminates on a fixed point, which can be used to implement non-Pauli operations. The exception to this rule is for certain pathological states, where the act of $z$-buffering causes the state to become long-range correlated, in that the renormalized correlation length $\tildeXi$ becomes infinite. This resource characterization is summarized in the following Theorem:

\begin{figure}[t]
  \centering
  \includegraphics[width=0.45\textwidth]{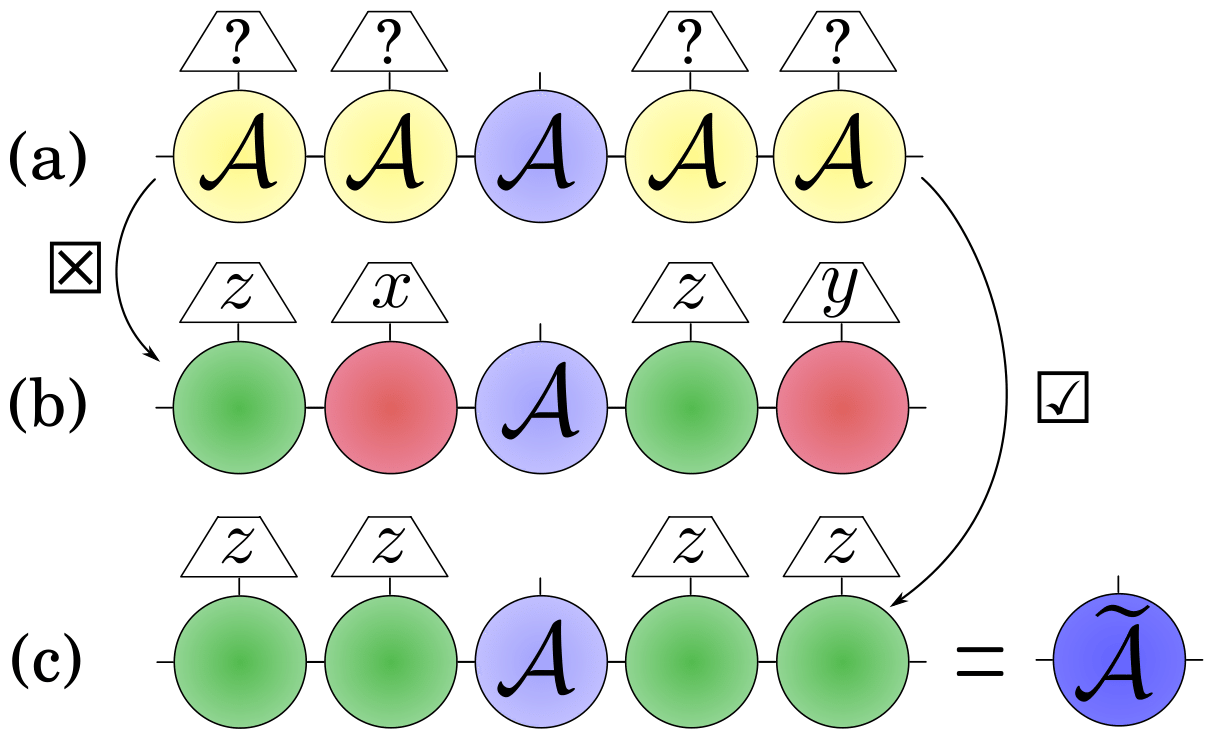}
  \caption{Schematic of renormalization procedure to manifest the quality of resource states. (a) To perform $z$-buffering, we choose a computational site, and measure $m$ surrounding sites in the Pauli basis. Here $m = 2$. (b) If our measurement fails to produce the all-$\ket{z}$ outcome, the computational site is measured in the Pauli basis, and we try again on another region. Since all of our measurement outcomes simply induce Pauli operations, the state of the protected space is (up to byproducts) unchanged. (c) If our measurement succeeds, the resource quality of our computational site is improved, at least when $\zeta_z$ is finite (Theorem~\ref{thm:ZB}).}
  \label{fig:Schematic}
\end{figure}
\begin{theorem}
  \label{thm:ZB}
  Consider any ground state of the 1D $S_4$ symmetry-protected topological ordered phase, which is characterized by a certain $z$-correlation length $\zeta_z$ and a renormalized correlation length $\tildeXi$. As long as $\zeta_z$ and $\tildeXi$ are both finite, the intrinsic entanglement of this state enables us to efficiently implement all one-qubit unitary operations under the setting of measurement-based quantum computation with arbitrarily high gate fidelity.
\end{theorem}
\begin{figure*}[th]
  \centering
  \includegraphics[width=1.0\textwidth]{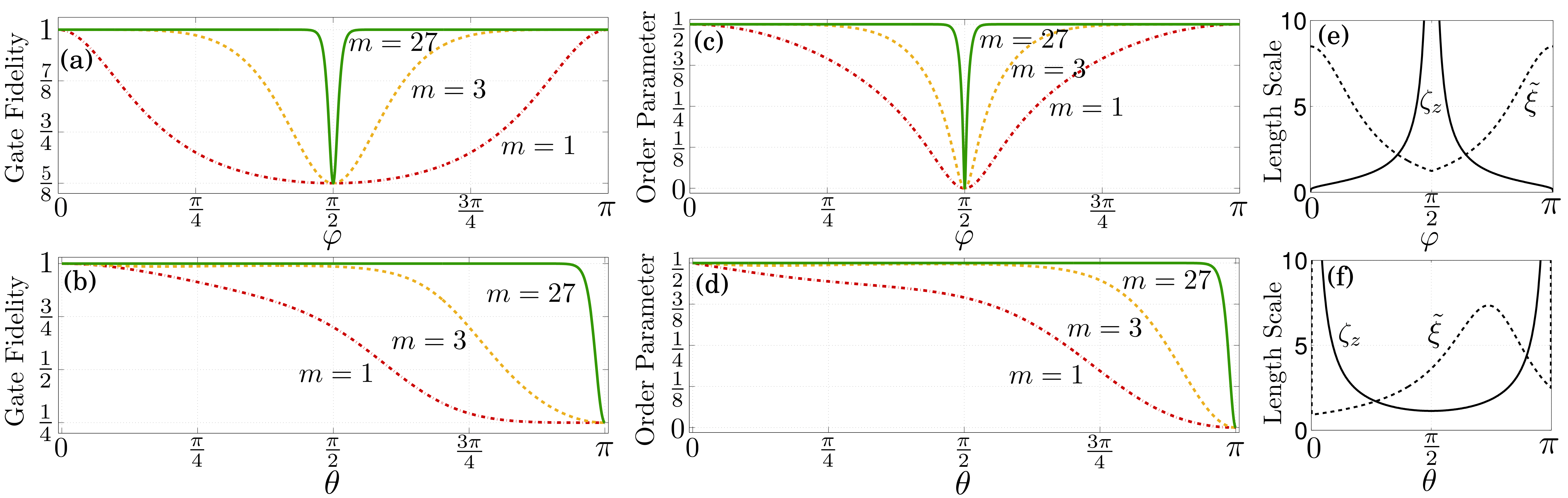}
  \caption{(a \& b) The gate fidelity for a protected space $\halfpi$ rotation about the $z$ axis, with resource states parameterized by $\varphi$, $\theta = \halfpi$ in (a), and by $\theta$, $\varphi = \frac{\pi}{4}$ in (b). The renormalized gate fidelity tends toward unity everywhere except at the regions of divergent $\zeta_z$, in agreement with Theorem~\ref{thm:ZB}. (c \& d) The renormalized order parameter $\sop{z}$ for the same set of parameters as in (a) and (b), respectively. The RG limit of $\sop{z}$ is $\onehalf$ everywhere that the RG limit of the gate fidelity is 1, in agreement with Theorem~\ref{thm:GF}. (e \& f) The $z$-correlation length and the renormalized correlation length, $\zeta_z$ and $\tildeXi$, shown for the same set of parameters as in (a) and (b), respectively. While both diverge at the poles of our parameter space, where our toy model is long-range correlated, the divergence of $\zeta_z$ at $\varphi = \halfpi$ is more surprising, and leads to a transition in the resource quality of our state there, as seen in (a).}
  \label{fig:six_plots}
\end{figure*}

The fact that our protocol enables the behavior described in Theorem~\ref{thm:ZB} is proven in Appendix~A. The main idea behind our proof \cite{Footnote2} is that our MPS resource state, by virtue of being in the $D_2$ SPTO phase, will have SPTO degeneracy in the protected space, but generally not in the junk space. When we postselect for a repeated $\ket{z}$ outcome, we maintain this protected space degeneracy, but preferentially amplify a one-dimensional subspace of the junk space. After enough buffering, the junk space is sufficiently restricted to this one-dimensional subspace, corresponding to the largest eigenvalue $\lambda_1$ of $a_z$, so that our renormalized system can be treated effectively like the AKLT state. The length scale over which this happens, $\zeta_z$, is set by the ratio of the largest to the second largest eigenvalue. The expected measurement overhead per gate required to achieve a gate fidelity $1 - \epsilon$ is
\begin{equation}
  \langle N \rangle = O \left( \zeta_z  \left( \frac{1}{\epsilon} \right)^{4 \zeta_z \log\lvert \frac{1}{\lambda_1} \rvert} 
  \log\left( \frac{1}{\epsilon} \right) \right).
\end{equation}
When the two largest eigenvalues of $a_z$ become degenerate, corresponding to a divergence in $\zeta_z$, $z$-buffering cannot completely restrict the junk space, and our RG flow stalls before reaching an AKLT-like state. 

Theorem~\ref{thm:ZB} says that the ground states of the $S_4$ SPTO phase generally share a common computational capability to implement perfect one-qubit gate operations. Since such capability is conveniently characterized in QIP by a measure called the gate fidelity, one could ask conversely ``Could the gate fidelity be utilized as an alternative, operationally-motivated order parameter for quantum phases of matter?''  Our second theorem below, proven in Appendix~B, states a surprising correspondence between the gate fidelity and (a type of) so-called string order parameter \cite{denNijs1989preroughening}, within the $S_4$ SPTO phase.
\begin{theorem}
  \label{thm:GF}
For any ground state in the 1D $S_4$ symmetry-protected topologically ordered phase with finite $\tildeXi$, the gate fidelity of all one-qubit operations in measurement-based quantum computation is perfect if and only if the order parameters $\sop{x}$ and $\sop{z}$ take maximal values of $\onehalf$ when these quantities are evaluated upon completion of renormalization. 
\end{theorem}
\noindent Note that our order parameters $\sop{x}$ and $\sop{z}$ are specializations of the string order parameters $R_\infty(u)$ from \cite{perez2008string} to the case of $\halfpi$ rotations about the $x$ and $z$ axes, $u_{r_x}$ and $u_{r_z}$. In \cite{perez2008string}, these string order parameters are argued to be capable of detecting the presence of quantum phase transitions between different SPTO phases. Our order parameters are given by:
\begin{equation}
  \label{eq:SOPDef}
  \sop{\mu} = \lim_{n \to \infty} \matrixel{\psi_\mu}{(u_{r_\mu})^{\otimes n} }{\psi_\mu}.
\end{equation}
The state $\ket{\psi_\mu}$ is the state of our many-body MPS after it has been mapped to the RG fixed point under $\mu$-buffering, where $\mu$ is either $x$ or $z$. While our bare spin chain possesses full $S_4$ symmetry, the process of renormalization breaks symmetry by picking out a preferred direction (the $x$ or $z$ axis). Consequently, the symmetry group of $\ket{\psi_\mu}$ is reduced to $D_4^{(\mu)}$, which consists of the 8 rotations within $S_4$ that preserve this preferred axis.

\emph{Illustration of Our Results.---}To demonstrate Theorems \ref{thm:ZB} and \ref{thm:GF}, we study the behavior of MPS's in the $S_4$ SPTO phase with a two-dimensional junk space. We have developed a general formalism based on representation theory \cite{miller2014upcoming}, and can show that spin-1 MPS's of this form make up a two-parameter family that is isomorphic to a sphere. Choosing variables $\theta$ and $\varphi$, with $0 \leq \theta < \pi$ and $0 \leq \varphi < 2\pi$, gives a unique parameterization of this family of MPS's. Because $S_4$ symmetry includes $D_2$ symmetry, these MPS's have well-defined protected and junk spaces, with component matrices $A_\mu(\theta, \varphi) = \sigma_\mu \otimes a_\mu(\theta, \varphi)$, and
\begin{equation}
  \label{eq:ToyModel}
  a_\mu(\theta, \varphi) = \frac{1}{\sqrt{3}} \left\{ \cos\left(\frac{\theta}{2}\right) I + e^{i \varphi} \sin\left(\frac{\theta}{2}\right) (\vec{n}_\mu \!\cdot\! \vec{\sigma}) \right\}.
\end{equation}
The Pauli-type operators $\vec{n}_\mu \!\cdot\! \vec{\sigma}$ form a triad defined by
\begin{equation}
  \label{eq:PauliTriad}
  -\onehalf \sigma_x + \frac{\sqrt{3}}{2} \sigma_y \; , \; -\onehalf \sigma_x - \frac{\sqrt{3}}{2} \sigma_y \; , \; \sigma_x \ ,
\end{equation}
for $\mu = x, y, z$ respectively. A numerical calculation of the gate fidelity, order parameter, and relevant length scales of states throughout the parameter space is shown in Figure~\ref{fig:six_plots}. We can see that the RG flow induced by $z$-buffering improves the gate fidelity of a $\halfpi$ rotation, an illustration by the ``most non-Pauli'' $z$-axis rotation, almost everywhere in our toy model. The points at which the gate fidelity is not improved are precisely those with divergent $\zeta_z$, in agreement with Theorem~\ref{thm:ZB}. Furthermore, we see remarkable similarity between the plots showing gate fidelity and those showing $\sop{z}$ in Figure~\ref{fig:six_plots}, both of which improve as the degree of $z$-buffering is increased. After sufficient renormalization (i.e., at $m=\infty$), the gate fidelity achieves its maximum value precisely when $\sop{z} = \sop{x} = \onehalf$, as stated in Theorem~\ref{thm:GF}.

There are a few singular states in our parameter space with regard to their behavior under renormalization. As shown in Figure~\ref{fig:six_plots}, the region with $\varphi = \pm \pi$ and any $\theta$, as well as the poles at $\theta = 0, \pi$, have divergent $\zeta_z$. This can be understood by noticing that $a_z$ is unitary at these points, so that $z$-buffering just acts as a change of basis on the junk space. Interestingly, the original correlation length $\xi$, does not diverge at $\varphi = \pm \pi$, so that this is a new kind of singular state only detected by our operationally motivated classification of quantum many-body states. In contrast, states at the poles ($\theta = 0, \pi$) are not within the $S_4$ SPTO phase, because the original MPS's are long-range correlated, having a divergent $\xi$. There is another singular state at $(\theta, \varphi) = (2 \arctan(2), 0)$, whose pathological behavior is discussed in Appendix~C.

\emph{Conclusion.---} We proved two theorems to demonstrate the intrinsic, quantum computational usefulness of the 1D $S_4$ SPTO phase as a ``universal'' quantum channel. We think that our physically feasible renormalization procedure, called $z$-buffering, is interesting on its own, because our state-insensitive protocol indicates that it is possible to harness such intrinsic capability of the phase without knowledge of microscopic details, at least as long as the states are guaranteed to be in the phase. As an outlook, since it is plausible that resource states for universal computation should generally possess such universal-channel capability in two or higher dimensions, our work is expected to serve as a stepping stone in the search for universal resource states in naturally-occurring quantum many-body systems. 

The work was supported in part by National Science Foundation grants PHY-1212445 and PHY-1314955.

\newpage
\appendix
\setcounter{theorem}{0}

In these Appendices, we give proofs to the two Theorems stated in the main text. For completeness, we restate them here.

\begin{theorem}
  Consider any ground state of the 1D $S_4$ symmetry-protected topological ordered phase, which is characterized by a certain $z$-correlation length $\zeta_z$ and a renormalized correlation length $\tildeXi$. As long as $\zeta_z$ and $\tildeXi$ are both finite, the intrinsic entanglement of this state enables us to efficiently implement all one-qubit unitary operations under the setting of measurement-based quantum computation with arbitrarily high gate fidelity.
\end{theorem}

\begin{theorem}
For any ground state in the 1D $S_4$ symmetry-protected topologically ordered phase with finite $\tildeXi$, the gate fidelity of all one-qubit operations in measurement-based quantum computation is perfect if and only if the order parameters $\sop{x}$ and $\sop{z}$ take maximal values of $\onehalf$ when these quantities are evaluated upon completion of renormalization. 
\end{theorem}

Before giving the proofs of these two Theorems, we introduce some facts and terminology useful for studying matrix product states.

While our original definition of MPS's consisted of a single-spin basis $\left\{ \ket{i} \right\}_{i=1}^d$ and a collection of matrices, $A_i$, one for each $\ket{i}$, we can treat these objects in a unified manner by defining a three-index MPS tensor, $\mathcal{A}$, as
\begin{equation}
  \label{eq:MPSComponent}
  \mathcal{A} = \sum_{i=1}^{d} A_i \ket{i}.
\end{equation}
The relevant Hilbert spaces here are the single-site space, referred to as the physical space, and the abstract Hilbert space which the $A_i$ act on, referred to as the virtual space. We refer to the operators $A_i$ as the component operators of $\mathcal{A}$.

Given a single-spin representation $u_G$ of a symmetry group $G$, the necessary and sufficient condition for a MPS to be invariant under this symmetry group is if our MPS tensor satisfies \cite{perez2008string, sanz2009matrix}
\begin{equation}
  \label{eq:PhysVirt}
  \sum\limits_{j=1}^d \left(u_G\right)_{i,j} A_j = U_G A_i U_G^\dagger.
\end{equation}
\noindent $U_G$ is generally allowed to be a projective representation, with $U_g U_h = e^{i \theta_{\!gh}} U_{gh}$, and the collection of $e^{i \theta_{\!gh}}$ is actually what determines a MPS's SPTO phase \cite{chen2011classification}. 

Finally, any MPS tensor can be put in a special canonical form \cite{perez2006matrix}, in which its component matrices satisfy the following relations:
\begin{eqnarray}
  \label{eq:CanonicalForm1}
  \mathcal{E}_I(I) &:=& \sum_{i=1}^d A_i I A_i^\dagger = I, \\
  \label{eq:CanonicalForm2}
  \mathcal{E}_I^\dagger(\Lambda) &:=& \sum_{i=1}^d A_i^\dagger \Lambda A_i = \Lambda,
\end{eqnarray}
\noindent where $I$ is the identity operator, and $\Lambda$ is a strictly positive operator satisfying $\tr(\Lambda) = 1$. Viewing Eqs.\ (\ref{eq:CanonicalForm1}) and (\ref{eq:CanonicalForm2}) as setting the largest eigenvalue of $\mathcal{E}_I$, the correlation length, $\xi$, of our state is determined by the magnitude of the second largest eigenvalue. If $I$ and $\Lambda$ are the only operators with eigenvalues of unit modulus, our MPS tensor is short-range correlated.

\section{Proof of Theorem~1}
\label{sec:ZBProof}
In this section, we give a proof of Theorem~1. For clarity, we first give a mathematical translation of each of the relevant terms in Theorem~1, for the case of $z$ rotations.
\begin{description}
  \item[Finite $\zeta_z$] $\zeta_z$ is set by the eigenvalues of $a_z$. When $a_z$ is a normal operator ($[a_z, a_z^\dagger] = 0$), $\zeta_z$ is defined in terms of the ratio of the largest and second largest eigenvalues of $a_z$, $\lambda_1$ and $\lambda_2$ respectively, as $\zeta_z = \left( -\log \left\lvert \frac{\lambda_2}{\lambda_1} \right\rvert \right)^{-1}$. In the case of non-normal $a_z$, the definition is the same, but $\lambda_1$ and $\lambda_2$ are required to be eigenvalues associated with distinct Jordan blocks, when $a_z$ is written in its Jordan normal form. The condition of finite $\zeta_z$ requires $\lvert \lambda_1 \rvert \neq \lvert \lambda_2 \rvert$.

  \item[Finite $\tildeXi$] $\tildeXi$ is defined in terms of the ratio of the largest and second largest eigenvalues of $\RgEi$, where $\RgEi$ is the quantum channel in \eqnref{eq:CanonicalForm1}, but with the matrices $A_i$ replaced by their renormalized counterparts $\tilde{A}_i$. We specify the action of the RG flow on the component matrices in \eqnref{eq:RGComps}, and show how this condition is needed near the end of our proof.

  \item[Gate Fidelity]  We quantify the fidelity of a single-qubit unitary operation by $F = \tr_P \{\tr_J [\mathcal{D}(\rho)] \ U_\Theta^{(P)} \rho^{(P)} U_\Theta^{(P)\dagger}\}$, where $\rho = \rho^{(P)} \otimes \rho^{(J)}$, and where $\mathcal{D}$ is the actual virtual space operation generated by measurement following renormalization. The choice of $\rho^{(P)}$ and $\rho^{(J)}$ isn't particularly important for our proof, but we will discuss their selection for Figures 2 and 3 at the end of Appendix~\ref{sec:ZBProof}.
\end{description}
Our proof involves modeling the MPS component operators under renormalization and showing that, given finite $\zeta_z$ and $\tildeXi$, the junk space components of the renormalized counterparts of $A_x$ and $A_y$ tend towards a common operator. In this case, we can perform a $z$-axis rotation using the same single-site measurement as for the AKLT state, and the gate fidelity of the resultant protected-space operation relative to the desired rotation will converge exponentially fast to unity.

In $z$-buffering, we postselect for obtaining the all-$\ket{z}$ outcome for $m$ sites on both sides of our computational site. The effect of this is to modify the MPS component operators of the computational site as follows:
\begin{equation}
  \label{eq:RGComps}
  A_\mu \mapsto \tilde{A}^{(m)}_\mu = (A_z^m) A_\mu (A_z^m) = \sigma_\mu \otimes \tilde{a}_\mu
\end{equation}
Since $A_z = \sigma_z \otimes a_z$ always has a trivial effect on the protected space, the interesting part of our proof involves looking at the iterated term $a_z^m$. If $a_z$ is a normal operator, then it can be diagonalized by expressing it in its eigenbasis. If $a_z$ is non-normal, then we can block diagonalize it by writing it in its Jordan canonical form. In this latter case,
\begin{equation}
  \label{eq:Jordan}
  a_z = \bigoplus\limits_{k=1}^p a_z^{(k)},
\end{equation}
\noindent where $a_z^{(k)} = \lambda_k I_{D_k} + Q_{D_k}$. Here, the index $k$ parameterizes the $p$ different Jordan blocks in the decomposition, each of which has dimension $D_k$. $I_{D_k}$ is the projector onto the $k$'th Jordan block and $Q_{D_k}$ is the operator whose matrix form has 1's immediately above the diagonal and 0's everywhere else. We assume that we have ordered the Jordan blocks by eigenvalue size, such that $\lvert\lambda_1\rvert \geq \lvert\lambda_2\rvert \geq \ldots \geq \lvert\lambda_p\rvert$. The form of \eqnref{eq:Jordan} includes the normal $a_z$ case as well (every Jordan block one-dimensional), so we only need to prove the efficacy of $z$-buffering for non-normal $a_z$.

Given this form, $a_z^m$ is
\begin{equation}
  \label{eq:junkExponentiated}
  a_z^m = (\lambda_1)^m \bigoplus\limits_{k=1}^p \left(\frac{\lambda_k}{\lambda_1}\right)^m P_k^{m},
\end{equation}
where $P_k = I_{D_k} + \lambda_k^{-1} Q_{D_k}$. If $\zeta_z$ is finite, then $\lambda_1$ is the unique largest eigenvalue, meaning that the weight attached to each $P_k^{m}$ with $k > 1$ decays exponentially with $m$ relative to that of $P_1^{m}$.

We now look at the element of $S_4$ corresponding to a $z$-axis rotation by $\halfpi$, whose physical space representation is denoted $u_{r_z}$. We know that $u_{r_z} \! \ket{z} = \ket{z}$, which in turn implies via \eqnref{eq:PhysVirt} that $[U_{r_z} , A_z] = 0$ on the virtual space. Representation theory (cf. \cite{miller2014upcoming}) can be used to show that the virtual symmetry operator $U_{r_z}$ decomposes along the protected-junk division as $U_{r_z} = U_{r_z}^{(P)} \otimes U_{r_z}^{(J)}$, with $U_{r_z}^{(P)} = e^{-i\frac{\pi}{4}\sigma_z}$ and $U_{r_z}^{(J)}$ satisfying $(U_{r_z}^{(J)})^2 = I$. This tells us that $U_{r_z}^{(J)}$ has eigenvalues of $\pm 1$, and the fact that $U_{r_z}$ and $A_z$ commute tells us that each Jordan block of \eqnref{eq:Jordan} can be labeled with one of these eigenvalues, denoted $\chi_k$. Symbolically,
\begin{equation}
  \label{eq:singleBlockCharacter}
  U_{r_z}^{(J)} a_z^{(k)} = \chi_k a_z^{(k)} .
\end{equation}
When restricted to the junk space, the condition \eqnref{eq:PhysVirt} becomes $U_{r_z}^{(J)} a_x U_{r_z}^{(J)\dagger} = a_y$, which also holds for the renormalized junk space components. If we define $\tilde{a}_\pm = \onehalf (\tilde{a}_x \pm \tilde{a}_y)$, this information, along with \eqnref{eq:junkExponentiated}, gives us
\begin{eqnarray}
  \label{eq:aPM}
  \tilde{a}_\pm &=& \onehalf (\tilde{a}_x \pm \tilde{a}_y) = \onehalf (\tilde{a}_x \pm U_{r_z}^{(J)} \tilde{a}_x U_{r_z}^{(J \dagger)}) \nonumber \\
                &=& \lambda_1^{2m} \sum\limits_{\substack{ j, k \\ \chi_j= \pm \chi_k}}  \left(\frac{\lambda_j \lambda_k}{\lambda_1^2}\right)^m P_j^{m} a_x P_k^{m},
\end{eqnarray}
where the condition $\chi_j= \pm \chi_k$ limits the range of summed indices in each case. In the large $m$ limit, the term associated with $\lambda_1^2$ dominates the sum in \eqnref{eq:aPM}. Since this term is contained within $\tilde{a}_+$, and not $\tilde{a}_-$, we see that $\tilde{a}_x$ and $\tilde{a}_y$ both converge to a common operator $\tilde{a}_+$ exponentially fast.

At any stage of renormalization, if we apply a projective measurement with measurement outcome $\ket{\psi_{z,\Theta}} = \cos\!\left(\frac{\Theta}{2}\right) \ket{x} - \sin\!\left(\frac{\Theta}{2}\right) \ket{y}$, the virtual operation implemented is
\begin{eqnarray}
  \label{eq:renormalizedMeasurement}
  A[\psi_{z,\Theta}] &=& \sigma_x \left[ \cos\!\left(\frac{\Theta}{2}\right) I \otimes \tilde{a}_x - i \sin\!\left(\frac{\Theta}{2}\right) \sigma_z \otimes \tilde{a}_y \right] \nonumber \\
                     &=& U_{+\Theta}^{(P)} \otimes \tilde{a}_+ + U_{-\Theta}^{(P)} \otimes \tilde{a}_- ,
\end{eqnarray}
where $U_{\pm\Theta}^{(P)} = \sigma_x \left[ \cos\!\left(\frac{\Theta}{2}\right) I \mp i \sin\!\left(\frac{\Theta}{2}\right) \sigma_z \right]$. The operation $\mathcal{D}$ used in our definition of gate fidelity is defined in terms of $A[\psi_{z,\Theta}]$ as $\mathcal{D}(\rho) = A[\psi_{z,\Theta}] \, \rho \, A[\psi_{z,\Theta}]^\dagger / \tr(A[\psi_{z,\Theta}] \, \rho \, A[\psi_{z,\Theta}]^\dagger)$. \eqnref{eq:renormalizedMeasurement} tells us that the operation induced by the measurement outcome is a coherent combination of a rotation by $\Theta$ with another rotation by $- \Theta$. The gate fidelity of the reduced operation on the protected space is set by the relative size of the junk space operators associated with the two rotations, and since $\tilde{a}_+$ is exponentially larger in norm than $\tilde{a}_-$ in the large $m$ limit, the gate fidelity between $A[\psi_{z,\Theta}]$ and $U_{+\Theta}^{(P)}$ will converge to unity exponentially fast.

In this proof, we explicitly required $S_4$ symmetry and finite $\zeta_z$ in order to give the description of $\tilde{a}_\pm$ in \eqnref{eq:aPM} and show the exponential separation between $\tilde{a}_+$ and $\tilde{a}_-$. Note, however, that our requirement of finite $\tildeXi$ was implicit in the assumption that $P_1^{m} a_x P_1^{m} \neq 0$. In Appendix~\ref{sec:RGXi}, we examine carefully a point in our toy model parameter space where this assumption is violated. Here we simply mention that for such a state, the renormalized junk space operator $\tilde{a}_z$ is exponentially larger in norm than both $\tilde{a}_x$ and $\tilde{a}_y$. For such a system, the renormalized identity-derived operator tends exponentially fast towards $\RgEi = \tilde{A}_z \odot \tilde{A}_z^\dagger$, which has degeneracy in the protected space portion of its eigenvalue spectrum. The correlation length of our system consequently increases exponentially with $m$, and this violates our assumption of finite $\tildeXi$. Thus, given our assumptions, we are guaranteed $P_1^{m} a_x P_1^{m} \neq 0$, and our proof of \mbox{Theorem~1} is complete.

We conclude with two remarks. First, in order to implement an arbitrary single qubit unitary gate to accuracy $\epsilon$, such that $F \geq 1 - \epsilon$, the expected overhead per gate, $\langle N \rangle$, is
\begin{equation}
  \langle N \rangle = O \left( \zeta_z  \left( \frac{1}{\epsilon} \right)^{4 \zeta_z \log\lvert \frac{1}{\lambda_1} \rvert} 
  \log\left( \frac{1}{\epsilon} \right) \right).
\end{equation}
This comes from our postselection success probability and gate fidelity having asymptotic scaling of $p_{succ} \sim \lvert \lambda_1 \rvert^{4m}$ and $F \sim 1 - e^{-\frac{m}{\zeta_z}}$. We note that for the case of non-normal $a_z$, the convergence of the operator $\tilde{a}_+$ to a definite limit form will generally happen at a rate that is polynomial, rather than exponential, in $m$. However, since we are only interested in applying $U_\Theta^{(P)}$ on the protected portion of our virtual space, and that is not hindered by any dynamics within the junk space, our measure of gate fidelity has been chosen to reflect only the reduced form of $A[\psi_{z,\Theta}]$ within the protected space. From \eqnref{eq:renormalizedMeasurement}, we see that this reduced form of $A[\psi_{z,\Theta}]$ converges to $U_\Theta^{(P)}$ at a rate that is exponential in $m$, regardless of the much slower convergence of $\tilde{a}_+$.

Second, we mention that although a specific choice for $\rho = \rho^{(P)} \otimes \rho^{(J)}$ is relatively unimportant for the scaling of our gate fidelity under renormalization, for the simulations involving our toy model we chose to use $\rho^{(P)} = \ketbra{+}{+}$ and $\rho^{(J)} = \onehalf I$, where $\ket{+}$ is the $+1$ eigenstate of $\sigma_x$. This choice of $\rho^{(P)}$ is natural for probing the fidelity of rotations about the $z$ axis, while the choice of $\rho^{(J)}$ corresponds to the limit of our junk space after sufficiently many unsuccessful postselection attempts. In particular, while unsuccessful postselection simply acts as identity channels (teleportation) on the protected space, at each stage we evolve the junk space by an unknown junk space operator, leading to an unknown final state which we take to be maximally mixed.

\section{Proof of Theorem~2}
\label{sec:GFProof}

To prove Theorem~2, we first have to give a definition of $\sop{\mu}$ that is more amenable to computation than that given in \eqnref{eq:SOPDef}. To this end, we define the channel $\mathcal{E}_{u_{r_\mu}}$ as the contraction of both indices of the physical symmetry $u_{r_\mu}$ with MPS tensors. Mathematically, $\mathcal{E}_{u_{r_\mu}} = \sum\limits_{\nu, \eta = 1}^3 \left( u_{r_\mu} \right)_{\nu,\eta} A_\eta \odot A_\nu^\dagger$. Using \eqnref{eq:PhysVirt}, we find that $\mathcal{E}_{u_{r_\mu}}$ is definable in terms of $\mathcal{E}_I$ as
\begin{equation}
  \mathcal{E}_{u_{r_\mu}} = U_{r_\mu} \circ \mathcal{E}_I \! \left( U_{r_\mu}^\dagger \odot \right).
\end{equation}
Now, using the standard method for calculating expectation values of tensor products of single-site operators on an MPS, we have that the string order parameters evaluated on the bare state are given by
\begin{eqnarray}
\label{eq:BareSOPDef}
  \mathcal{O}_{D_4}^{(\mu)} &=& \lim_{n \to \infty} \tr\!\left( \Lambda \circ (\mathcal{E}_{u_{r_\mu}})^{n} [I] \right) \nonumber \\
                &=& \lim_{n \to \infty} \tr\!\left( \Lambda U_{r_\mu} (\mathcal{E}_I)^{n} [U_{r_\mu}^\dagger] \right).
\end{eqnarray}
$(\mathcal{E}_{W})^{n}$ here means the $n$-fold iterated operation of the quantum channel $\mathcal{E}_{W}$ ($W$ representing either $u_{r_\mu}$ or $I$), and $\Lambda$ denotes the left limit edge mode of our MPS, defined implicitly in \eqnref{eq:CanonicalForm2}. The value of $\sop{\mu}$, the renormalized string order parameters, are given by the same expression as in \eqnref{eq:BareSOPDef}, but with $\RgEi$ in place of $\mathcal{E}_I$.

The action of $z$-buffering on the virtual space is described in Appendix~\ref{sec:ZBProof}. For our purposes here, we only add that the rescaling in the normalization of our state that is required by postselection can be compactly expressed as the requirement that the spectral radius of $\RgEi$ is 1. The results of \cite{perez2006matrix}, together with the assumption that our renormalized MPS tensor is short-range correlated (finite $\tildeXi$), then tell us that we can pick a basis for our junk space which puts our renormalized MPS tensor in canonical form. In this form, our channel $\RgEi$ satisfies the following conditions:
\begin{eqnarray}
  \label{eq:NewCanonicalForm1}
  \RgEi(I_P \otimes \Pi_z) &=& \sum_{\mu=1}^3 \tilde{A}_\mu (I_P \otimes \Pi_z) \tilde{A}_\mu^\dagger = I_P \otimes \Pi_z, \\
  \label{eq:NewCanonicalForm2}
  \RgEi^\dagger(I_P \otimes \tilde{\Lambda}) &=& \sum_{\mu=1}^3 \tilde{A}_\mu^\dagger (I_P \otimes \tilde{\Lambda}) \tilde{A}_\mu = I_P \otimes \tilde{\Lambda},
\end{eqnarray}
\noindent where $\Pi_z$ is a projector onto the section of our junk space with non-vanishing support at the RG fixed point, and $\tilde{\Lambda}$ is a strictly positive operator of unit trace, whose support is exactly $\Pi_z$.

Our proof of Theorem~2 consists of a case-by-case analysis of the renormalized junk space operators $\tilde{a}_x$ and $\tilde{a}_y$, depending on whether or not $\sop{z} = \onehalf$. We first show that if $\sop{z} = \onehalf$, then $\tilde{a}_x = \tilde{a}_y$ at the RG fixed point. In this case, a $z$-axis rotation implemented using the $z$-buffered MPS will have perfect gate fidelity. We then show that if $\sop{z} \neq \onehalf$, then $\tilde{a}_x \neq \tilde{a}_y$. This causes the gate fidelity of our attempted $z$-axis rotation to be less than unity at the RG limit. Proving both of these implications under the assumption of finite $\tildeXi$ suffices to prove Theorem~2.

For the first direction of the proof, we note that finite $\tildeXi$ means that $I_P \otimes \Pi_z$ and $I_P \otimes \tilde{\Lambda}$ are the only fixed points of Eqs. (\ref{eq:NewCanonicalForm1}) and (\ref{eq:NewCanonicalForm2}). This, together with the results of \cite{novotny2012asymptotic}, tells us that $(\RgEi)^{n}$ has the limit form
\begin{equation}
  \label{eq:LimitCanonical}
  \lim_{n \to \infty} (\RgEi)^{n} = \onehalf \tr[(I_P \otimes \tilde{\Lambda}) \odot] \ I_P \otimes \Pi_z.
\end{equation}
We now insert the form of \eqnref{eq:LimitCanonical} into (the renormalized counterpart of) \eqnref{eq:BareSOPDef} to get
\begin{eqnarray}
  \label{eq:OrderParameter}
  \sop{z} &=& \frac{1}{4} \tr[(I_P \otimes \tilde{\Lambda}) U_{r_z}] \tr[(I_P \otimes \tilde{\Lambda}) U_{r_z}^\dagger] \nonumber \\
  &=& \frac{1}{4} \left\lvert \tr((I_P \otimes \tilde{\Lambda}) U_{r_z}) \right\rvert^2 .
\end{eqnarray}
As mentioned in Appendix~\ref{sec:ZBProof}, the virtual unitary $U_{r_z}$ decomposes as $U_{r_z} = U_{r_z}^{(P)} \otimes U_{r_z}^{(J)}$. Furthermore, $\left\lvert \tr \big( U_{r_z}^{(P)} \big) \right\rvert^2 = 2$ for all states in the $D_2$ SPTO phase, so the value of $\sop{z}$ only depends on the behavior of the junk space.

To figure out this value, we first define $\tilde{U}_{r_z}$ to be the restriction of $U_{r_z}^{(J)}$ to the support of the junk space at the RG fixed point, $\tilde{U}_{r_z} := \Pi_z U_{r_z}^{(J)} \Pi_z$. The fact that $(U_{r_z}^{(J)})^2 = I_J$, along with $[ U_{r_z}^{(J)}, a_z] = 0$, shows that $(\tilde{U}_{r_z})^2 = \Pi_z$. Consequently, $\tilde{U}_{r_z}^{(J)}$ has eigenvalues of $\pm 1$, and we can write it as
\begin{equation}
  \label{eq:UrzRenormalized}
  \tilde{U}_{r_z}^{(J)} = \Pi_z^{+} - \Pi_z^{-},
\end{equation}
for projectors $\Pi_z^{+}$ and $\Pi_z^{-}$, which satisfy $\Pi_z^{+} \Pi_z^{-} = 0$ and $\Pi_z^{+} + \Pi_z^{-} = \Pi_z$.

Feeding this information into \eqnref{eq:OrderParameter} gives
\begin{equation}
  \label{eq:OrderParameter2}
  \sop{z} = \frac{1}{2} \left\lvert \tr(\tilde{\Lambda} \tilde{U}_{r_z}^{(J)}) \right\rvert^2 = \onehalf \left\lvert \tr[ \tilde{\Lambda} (\Pi_z^{+} - \Pi_z^{-}) ] \right\rvert^2.
\end{equation}
Now, we come to the meat of our proof. Since $\tilde{\Lambda}$ is a strictly positive operator with unit trace, the only way to have $\sop{z} = \onehalf$ is to have either $\Pi_z^+ = 0$ or $\Pi_z^- = 0$. In this case, $\tilde{U}_{r_z}^{(J)} = \pm \Pi_z$, which says our $z$-axis rotation acts trivially on the junk space of our renormalized MPS tensor.
\begin{figure*}[th]
  \centering
  \includegraphics[width=1.0\textwidth]{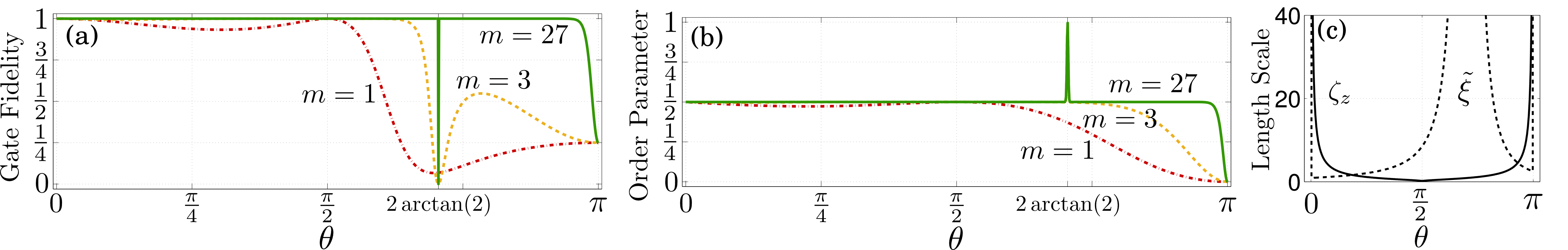}
  \caption{(a) The gate fidelity of a protected space $\halfpi$ rotation about the $z$ axis, for resource states along a North-to-South traversal of our parameter space ($\theta$ variable, $\varphi = 0$). While the renormalized gate fidelity tends toward unity most everywhere, it stays well below 1 at the South pole and at $\theta_c = 2 \arctan(2)$, where $\tildeXi$ diverges. (b) The renormalized order parameter $\sop{z}$ for the same set of parameters as in (a). The value of $\sop{z}$ is $\onehalf$ everywhere except at the South pole and, more surprisingly, at $\theta_c$. This unintuitive behavior can be explained by the divergence of $\tildeXi$. (c) The length scales $\zeta_z$ and $\tildeXi$ for the same parameters. Both quantities diverge at the poles, but the divergence of $\tildeXi$ at $\theta_c$ leads to the unexpected behavior seen in (a) and (b).}
  \label{fig:three_plots}
\end{figure*}
This last fact, which comes from assuming $\sop{z} = \onehalf$, lets us prove the equality of $\tilde{a}_x$ and $\tilde{a}_y$. This follows because
\begin{equation}
  \label{eq:JunkSpaceEquality}
  \tilde{a}_x = \tilde{U}_{r_z}^{(J)} \tilde{a}_y \tilde{U}_{r_z}^{(J)\dagger} = (\pm \Pi_z) \tilde{a}_y (\pm \Pi_z) = \tilde{a}_y.
\end{equation}
This gives the first direction of our proof.

For the other direction, assume that $\sop{z} \neq \onehalf$. In this case, \eqnref{eq:OrderParameter2} tells us that $\Pi_z^+$ and $\Pi_z^-$ are both non-zero. Thus, $\tilde{U}_{r_z}^{(J)}$ is not simply $\pm \Pi_z$. What does this say about $\tilde{a}_x$ and $\tilde{a}_y$? We can answer this by looking at the commutator $[\tilde{U}_{r_z}^{(J)}, \tilde{a}_x]$. If this is non-zero, then $\tilde{a}_x \neq \tilde{a}_y$. From \eqnref{eq:aPM}, we see that the norms of $\tilde{a}_+$ and $\tilde{a}_-$ do not become exponentially separated in the RG limit, and thus our renormalized state cannot be used to implement high-fidelity $z$-axis rotations on the protected space.

On the other hand, if $[\tilde{U}_{r_z}^{(J)}, \tilde{a}_x] = 0$, then we can take linear combinations of this with the commutator $[\Pi_z, \tilde{a}_x]$, which is always zero, to obtain
\begin{eqnarray}
	\lbrack \Pi_z^{+}, \tilde{a}_x \rbrack &=& 0 \\
	\lbrack \Pi_z^{-}, \tilde{a}_x \rbrack &=& 0,	
\end{eqnarray}
and the same for $\tilde{a}_y$. Since additionally, $[\Pi_z^{\pm}, \tilde{a}_z] = 0$ always, these facts together tell us that $\RgEi$ has two independent fixed points, $I_P \otimes \Pi_z^{+}$ and $I_P \otimes \Pi_z^{-}$. But this contradicts the assumption of finite $\tildeXi$, and thus cannot be the case for our system. Thus, we must have $\tilde{a}_x \neq \tilde{a}_y$, which completes the second desired implication, and thus finishes our proof of Theorem~2.

\section{The Renormalized Correlation Length}
\label{sec:RGXi}

While the physical interpretation of $\zeta_z$ in our protocol is straightforward, simply being the characteristic length scale of our RG flow, the interpretation of $\tildeXi$ is somewhat less clear. In this section, we take a closer look at this quantity by means of our toy model. Our toy model has three points for which $\tildeXi$ is divergent. Two of these points, those on the poles of our parameter space, start out as long-range correlated states before $z$-buffering, and thus aren't particularly interesting. However, the last point, lying at $(\theta, \varphi) = (\theta_c, 0)$ (for $\theta_c := 2 \arctan(2)$), possesses a correlation length that is only made divergent under $z$-buffering. We hope to clarify this behavior here by exhibiting the somewhat pathological behavior of this point under $z$-buffering.

Since the component matrices of our toy model are normal, the proof of Appendix~\ref{sec:ZBProof} simplifies considerably, and can be phrased as follows:
\begin{itemize}
  \item Z-buffering acts as $A_\mu \mapsto \tilde{A}^{(m)}_\mu = (A_z^m) A_\mu (A_z^m)$, which has a non-trivial effect only on the junk space.
  \item Since the eigenvector of $a_z$ with largest eigenvalue is either $\ket{+}$ (when $\real(e^{i\varphi}) > 0$) or $\ket{-}$ (when $\real(e^{i\varphi}) < 0$), we have $\lim\limits_{m \to \infty} (a_z)^m \sim \ketbra{\pm}{\pm}$.
  \item Thus, for the portion of our parameter space with $-\halfpi < \varphi < \halfpi$, our junk space components at the $m\!\to\!\infty$ limit satisfy $\tilde{a}_x = \tilde{a}_y = [\cos\big(\frac{\theta}{2}\big) - \onehalf e^{i\varphi}\sin\big(\frac{\theta}{2}\big)] \ketbra{+}{+}$, and $\tilde{a}_z = [\cos\big(\frac{\theta}{2}\big) + e^{i\varphi}\sin\big(\frac{\theta}{2}\big)] \ketbra{+}{+}$.
\end{itemize}
However, setting $(\theta, \varphi)$ to $(\theta_c, 0)$ shows that at this point, $\tilde{a}_x = \tilde{a}_y = 0$. In the language of Appendix~\ref{sec:ZBProof}, this is equivalent to $P_1 a_x P_1 = 0$, which is to say that the leading order term in $\tilde{a}_+$ vanishes. In this case, from \eqnref{eq:aPM} and from the fact that $U_{r_z}^{(J)} = \sigma_x$ for our toy model, we see that the dominant terms in our junk space components lie within $\tilde{a}_-$. This conclusion, along with \eqnref{eq:renormalizedMeasurement}, tells us that the RG limit of our effective protected space operation is $U_{-\Theta}^{(P)}$, a rotation in the opposite direction than we intended. While this can be accounted for by changing the interpretation we attach to our measurement outcomes, this selective change in interpretation would render our protocol no longer state-insensitive. Thus, for consistency, we must rule this state out as a valid resource state for MQC under $z$-buffering. The sharp dip in the gate fidelity seen at $\theta_c$ in Figure~\ref{fig:three_plots} is the natural consequence of making such a consistent choice of gate fidelity.

Finally, we explain the strange behavior seen in the value of $\sop{z}$ at $(\theta_c, 0)$. While this behavior appears quite surprising, it is explained by the fact that the channel $\RgEi$ is $\tilde{A}_z \odot \tilde{A}_z^\dagger$ at this point. Thus, even though the junk space of our system is restricted to a one-dimensional subspace here, the protected space portion of $\RgEi$ becomes degenerate at the RG fixed point. Consequently, the limit form of $\sop{z}$ is not given by \eqnref{eq:OrderParameter}, but rather by
\begin{eqnarray}
  \label{eq:OrderParameterCrit}
  \sop{z} &=& \frac{1}{2} \tr(U_{r_z}^{(P)} U_{r_z}^{(P)\dagger}) \tr(\tilde{\Lambda} U_{r_z}^{(J)}) \tr(\tilde{\Lambda} U_{r_z}^{(J)\dagger}) \nonumber \\
          &=& \left\lvert \tr(\tilde{U}_{r_z}^{(J)}) \right\rvert^2 = 1.
\end{eqnarray}
This completes our examination of the behavior of the $(\theta, \varphi) = (\theta_c, 0)$ point in our parameter space. Our intent in this, besides simply giving a complete account of our toy model, is to demonstrate that states with divergent $\tildeXi$ have rather pathological behavior that makes them unfit for use in our protocol. Thus, even without a concrete physical interpretation for this quantity, the stipulation of finite $\tildeXi$ is clearly necessary in both of our theorems.


\begin{thebibliography}{99}

\bibitem{li2008entanglement}
	H. Li and F. D. M. Haldane,
	Phys. Rev. Lett. \textbf{101}, 010504 (2008).

\bibitem{pollmann2010entanglement}
	F. Pollmann, A. M. Turner, E. Berg, and M. Oshikawa,
	Phys. Rev. B \textbf{81}, 064439 (2010).

\bibitem{chen2010local}
	X. Chen, Z.-C. Gu, and X.-G. Wen,
	Phys. Rev. B \textbf{82}, 155138 (2010).

\bibitem{deChiara2012entanglement}
	G. De Chiara, L. Lepori, M. Lewenstein, and A. Sanpera,
	Phys. Rev. Lett. \textbf{109}, 237208 (2012).
	
\bibitem{raussendorf2001one}
	R. Raussendorf and H. J. Briegel,
	Phys. Rev. Lett. \textbf{86}, 5188 (2001).
	
\bibitem{raussendorf2003measurement}
	R. Raussendorf, D. E. Browne, and H. J. Briegel,
	Phys. Rev. A \textbf{68}, 022312 (2003).
	
\bibitem{vidal2003efficient}
	G. Vidal,
	Phys. Rev. Lett. \textbf{91}, 147902 (2003).

\bibitem{cirac2012goals}
	J. I. Cirac and P. Zoller
	Nat. Phys. \textbf{8}, 264 (2012).

\bibitem{korenblit2012quantum}
	S. Korenblit \emph{et al},
	New J. Phys. \textbf{14}, 095024 (2012).

\bibitem{georgescu2014quantum}
	I. M. Georgescu, S. Ashhab, and F. Nori,
	Rev. Mod. Phys. \textbf{86}, 153 (2014).
	
\bibitem{eisert2014quantum}
	J. Eisert, M. Friesdorf, and C. Gogolin,
	Nat. Phys. \textbf{11}, 124 (2015).

\bibitem{cohen2014proposal}
	I. Cohen and A. Retzker,
	Phys. Rev. Lett. \textbf{112}, 040503 (2014).

\bibitem{wen2007quantum}
	X.-G. Wen,
	\emph{Quantum Field Theory of Many-body Systems}
	(Oxford University Press, Oxford, 2007).

\bibitem{hasan2010topological}
	M. Z. Hasan and C. L. Kane,
	Rev. Mod. Phys. \textbf{82}, 3045 (2010).
	
\bibitem{qi2011topological}
	X.-L. Qi and S.-C. Zhang,
	Rev. Mod. Phys. \textbf{83}, 1057 (2011).

\bibitem{kitaev2003fault}
	A. Kitaev,
	Ann. Phys. \textbf{303}, 2 (2003).

\bibitem{kitaev2010topological}
	A. Kitaev and C. Laumann,
	\emph{Topological phases and quantum computation}
	(Oxford University Press, Oxford, 2010), pp. 101–125.
  
\bibitem{briegel2001persistent}
	H. J. Briegel and R. Raussendorf,
	Phys. Rev. Lett. \textbf{86}, 910 (2001).
  
\bibitem{verstraete2004valence}
	F. Verstraete and J. I. Cirac,
	Phys. Rev. A \textbf{70}, 060302 (2004).
	
\bibitem{gross2007novel}
	D. Gross and J. Eisert,
	Phys. Rev. Lett. \textbf{98}, 220503 (2007).
	
\bibitem{gross2007measurement}
	D. Gross, J. Eisert, N. Schuch, and D. Pérez-García,
	Phys. Rev. A \textbf{76}, 052315 (2007).
	
\bibitem{cai2010universal}
	J.-M. Cai, A. Miyake, W. Dür, H. J. Briegel,
	Phys. Rev. A \textbf{82}, 052309 (2010).
	
\bibitem{miyake2011quantum}
	A. Miyake,
	Ann. Phys. \textbf{326}, 1656 (2011).
	
\bibitem{wei2011affleck}
	T.-C. Wei, I. Affleck, and R. Raussendorf,
	Phys. Rev. Lett. \textbf{106}, 070501 (2011).
	
\bibitem{li2011thermal}
	Y. Li \emph{et al},
	Phys. Rev. Lett. \textbf{107}, 060501 (2011).

\bibitem{browne2008phase}
	D. E. Brown, M. B. Elliott, S. T. Flammia, S. T. Merkel, A. Miyake, and A. J. Short,
  	New. J. Phys. \textbf{10}, 023010 (2008).
  	
\bibitem{doherty2009identifying}
	A. C. Doherty and S. D. Bartlett,
  	Phys. Rev. Lett. \textbf{103}, 020506 (2009).
  	
  \bibitem{barrett2009transitions}
	S. D. Barrett, S. D. Bartlett, A. C. Doherty, D. Jennings, and T. Rudolph,
	Phys. Rev. A \textbf{80}, 062328 (2009).
	
\bibitem{skrovseth2009phase}
	S. O. Skrøvseth and S. D. Bartlett,
	Phys. Rev. A \textbf{80}, 022316 (2009).
	
\bibitem{bartlett2010quantum}
	S. D. Bartlett, G. K. Brennen, A. Miyake and, J. M. Renes,
	Phys. Rev. Lett. \textbf{105}, 110502 (2010).

\bibitem{darmawan2012measurement}
	A. S. Darmawan, G. K. Brennen, and S. D. Bartlett,
	New J. Phys. \textbf{14}, 013023 (2012).

\bibitem{else2012symmetrynewjphys}
	D. V. Else, S. D. Bartlett, and A. C. Doherty,
	New J. Phys. \textbf{14}, 113016 (2012).

\bibitem{fujii2012topologically}
	K. Fujii and T. Morimae,
	Phys. Rev. A \textbf{85}, 010304(R) (2012).

\bibitem{fujii2013measurement}
	K. Fujii, Y. Nakata, M. Ohzeki, and M. Murao,
	Phys. Rev. Lett. \textbf{110}, 120502 (2013).

\bibitem{gu2009tensor}
	Z. C. Gu and X. G. Wen,
	Phys. Rev. B \textbf{80}, 155131 (2009).

\bibitem{chen2011classification}
	X. Chen, Z. C. Gu, and X. G. Wen,
	Phys. Rev. B \textbf{83}, 035107 (2011).
	
\bibitem{schuch2011classifying}
	N. Schuch, D. Pérez-García, and J. I. Cirac,
	Phys. Rev. B \textbf{84}, 165139 (2011).

\bibitem{pollmann2012symmetry}
	F. Pollmann, E. Berg, A. M. Turner, and M. Oshikawa,
	Phys. Rev. B \textbf{85}, 075125 (2012).
	
\bibitem{bravyi2005universal}
	S. Bravyi and A. Kitaev,
	Phys. Rev. A \textbf{71}, 022316 (2005).

\bibitem{fannes1992finitely}
	M. Fannes, B. Nachtergaele, and R. F. Werner,
	Commun. Math. Phys. \textbf{144}, 443 (1992)

\bibitem{perez2006matrix}
	D. Pérez-García, F. Verstraete, M. M. Wolf, and J. I. Cirac,
	Quantum Inf. Comput. \textbf{7}, 401 (2007).

\bibitem{affleck1988valence}
	I. Affleck, T. Kennedy, E. H. Lieb, and H. Tasaki,
	Comm. Math. Phys. \textbf{115}, 477 (1988).

\bibitem{brennen2008measurement}
	G. K. Brennen and A. Miyake,
	Phys. Rev. Lett. \textbf{101}, 010502 (2008).

\bibitem{else2012symmetryprl}
	D. V. Else, I. Schwarz, S. D. Bartlett, and A. C. Doherty,
	Phys. Rev. Lett. \textbf{108}, 240505 (2012).

\bibitem{Footnote1}
	Their actual result is more general, but it is only the spin-1 $D_2$ version of their result that is necessary for what follows.

\bibitem{popp2005analytic}
	M. Popp, F. Verstraete, M. A. Martín-Delgado, and J. I. Cirac,
	Phys. Rev. A \textbf{71}, 042306 (2005).

\bibitem{venuti2005analytic}
	L. Campos Venuti and M. Roncaglia,
	Phys. Rev. Lett. \textbf{94}, 207207 (2005).

\bibitem{Footnote2}
    While the intuition is essentially the same, the case of non-normal $a_z$  (i.e., $[a_z, a_z^\dagger] \neq 0$) needs some elaboration.

\bibitem{denNijs1989preroughening}
	M. den Nijs and K. Rommelse,
	Phys. Rev.B \textbf{40}, 4709 (1989).

\bibitem{perez2008string}
	D. Pérez-García, M. M. Wolf, M. Sanz, F. Verstraete, and J. I. Cirac,
	Phys. Rev. Lett. \textbf{100}, 167202 (2008).

\bibitem{miller2014upcoming}
	J. Miller and A. Miyake (to be published).
	
\bibitem{sanz2009matrix}
	M. Sanz, M. M. Wolf, D. Pérez-García, and J. I. Cirac,
	Phys. Rev. A \textbf{79}, 042308 (2009).

\bibitem{novotny2012asymptotic}
	J. Novotný, G. Alber, and I. Jex,
	J. Phys. A: Math. Theor. \textbf{42}, 282003 (2009).

\end{thebibliography}
\end{document}